\documentclass[preprint,amsmath,amssymb]{revtex4}
\usepackage{graphicx}
\usepackage{tabularx}
\usepackage{makecell}
\usepackage{dcolumn}
\usepackage{bm}
\DeclareMathOperator{\sech}{sech}
\begin{document}
\title{Stationary analogs of de Sitter and anti-de Sitter spacetimes: Evidence in favor of $\Lambda$ as a dark fluid}
\author{${\rm M.\;Nouri}$-${\rm Zonoz}^{\;(a)}$\footnote{Electronic
address:~nouri@ut.ac.ir \; (Corresponding author)} and ${\rm A.\;Nouri}$-${\rm Zonoz}^{\;(b)}$\footnote {Electronic
address:~ahmad.nouri.zonn@gmail.com } }
\affiliation{(a): Department of Physics, University of Tehran, North Karegar Ave., Tehran 14395-547, Iran.\\ 
(b): Department of Physics, Shahid Beheshti University, G.C., Evin, Tehran 19839, Iran.}
\begin{abstract}
We introduce novel Einstein spaces which are the {\it stationary analogs of de Sitter and ani-de Sitter} spacetimes. 
Having $\Lambda$ as their only parameter, the inherent anisotropy in these solutions appears as a dilemma if we treat 
the cosmological term as a constant background curvature.
Since the cosmological term is an inseparable element of the current cosmological models,
further development of these models  as well as resolution of the problems related to this term, including any quantum aspect 
will crucially depend on the true nature of this term. These solutions, which to the best of our knowledge  
appear here for the first time in the literature, not only open a new direction to study the cosmological term, but also  
provide strong evidence in favor of the cosmological term as a dark fluid, hence furnishing a resolution to 
the above mentioned dilemma.
\end{abstract}

\maketitle
\section{Introduction and motivation}
Since its introduction into the GR field equations by Einstein in 1917, the  question of the nature of the cosmological term  and its role in different cosmological 
models have always been of high interest among theoretical physicists as well as cosmologists \cite{Weinberg, Witten, Sahni}.
Whether it is a  constant background curvature of a geometric nature in the left 
hand side of the Einstein modified field equations (EMFE) or a matter source of a perfect fluid nature in the right hand side of the Einstein field 
equations (EFE), could obviously have important implications in further development of cosmological models.\\
Treating the cosmological term as a dark fluid component in the right hand side of the Einstein field equations (EFE), it was shown that the so called de Sitter-type 
spacetimes are the only {\it static  single-component} perfect fluid solutions of the Einstein field equations in the {\it non-comoving} frames \cite{NKR}. They include the 
well known de sitter (dS) and anti-de Sitter (AdS) spaces and their axially and cylindrically symmetric counterparts. They are all static Einstein spaces, with $\Lambda$ 
as their only parameter, which go over smoothly into the Minkowski spacetime in the absence of the cosmological constant.
Characterizing them in this way, the authors in \cite{NKR} have resolved the apparent paradox raised by some authors \cite{Rindler, GP}, 
on how one should interpret the anisotropic expansions of axially and cylindrically symmetric de Sitter-type spacetimes.\\
Indeed that is why, at first sight, they were expected to be just the good old dS space, only in different coordinate systems.
An expectation which was not fulfilled  when their curvature invariants as well as their dynamical forms in the 
{\it comoving synchronous} coordinate systems, showed clearly that they are genuinely different spacetimes. 
These findings motivated the idea that one should consider a {\it perfect fluid nature} for the cosmological term and assign a 4-velocity to this dark fluid, 
so that to be able to interpret and justify their {\it anisotropic expansions} \cite{NKR}.\\
Here we extend the above ideas by introducing stationary analogs of dS and AdS spacetimes, 
and then argue their relevance to the nature of the cosmological term. 
\subsection{Stationary analogs of de Sitter and anti-de Sitter spacetimes}
In this section we introduce explicit forms for {\it stationary} cylindrically symmetric {\it Einstein spaces} with positive and negative cosmological constants which reduce 
to the Minkowski spacetime in the absence of the cosmological constant \footnote{It is noticed that not every Einstein space has such a characteristic, for example Kottler 
space reduces to the Schwarzschild spacetime in the absence of the cosmological constant.}.
In other words $\Lambda$ is the only parameter of these solutions, and so one could call them the stationary analogs of dS and AdS spacetimes.
They are equivalent to the stationary cylindrically symmetric solutions of the Einstein field equations with a one-component perfect fluid source with the 
equation of state $p = - \rho = constant$.
It is known that stationary cylindrically symmetric solutions have the following general form \cite{Exact1}
\begin{equation}\label{ds06}
d{s^2} = F(\rho) ({dt} + A(\rho)d\phi)^2 - d\rho^2 - e^{2K(\rho)} dz^2 - G(\rho) d\phi^2,
\end{equation}
which is already written in the so called $1+3$ form with the $A(\rho)$ playing the role of a gravitomagnetic potential, from which the 
gravitomagnetic field is calculated \cite{MN}. 
The general form of stationary cylindrically symmetric Einstein spaces for both positive and negative cosmological constant have already been introduced 
\cite{Krasinski, Santos, Mac} in the literature \footnote{For a recent review on cylindrical gravitational systems refer to \cite{Bro}.}. 
In particular the following useful form for such general solutions were introduced in \cite{Mac},
\begin{align}\label{ds007}
ds^2 &= G_{0}^{2/3}\bigg([T_{1}^2 e^{2(p_{0}-1/3)U} - X_{1}^2 e^{2(p_{3}-1/3)U}]dt^2 - Z^2 e^{2(p_{2} - 1/3)U}dz^2 \nonumber \\&- [X_{2}^2 e^{2(p_{3}-1/3)U} - 
T_{2}^2 e^{2(p_{0}-1/3)U}]d{\phi}^2 \nonumber \\& - 
2[X_{1}X_{2} e^{2(p_{3}-1/3)U} - T_{1}T_{2} e^{2(p_{0}-1/3)U}]dt d\phi \bigg) - d\rho^2,  
\end{align}
in which $Z,\; X_1,\; X_2,\; T_1,\; T_2,\; p_0,\; p_2\; {\rm and}\; p_3$ are constants and functions $G_{0}(\rho)$ and $U(\rho)$ are given by,
\footnote{Indeed, in the case of $\Lambda < 0$, there are three 
different choices for each of these functions, but we have chosen the one which is the direct counterpart of the solution with $\Lambda > 0$.}
\begin{gather}
G_{0}(\rho) = \frac{1}{\sqrt{3\Lambda}}\sin(\sqrt{3\Lambda}\rho), \;\;\;\;\;\;\; \Lambda > 0 \\
G_{0}(\rho) = \frac{1}{\sqrt{-3\Lambda}}\sinh (\sqrt{-3\Lambda}\rho) \;\;\;\;\;\;\; \Lambda < 0 \\
U(\rho)= \ln \;\tan\frac{1}{2}(\sqrt{3\Lambda}\rho) \;\;\;\;\;\;\;\;\; \Lambda > 0 \\
U(\rho) = \ln \;\tanh\frac{1}{2}(\sqrt{-3\Lambda}\rho) \;\;\;\;\;\;\;\; \Lambda < 0,
\end{gather}
also the three constants $ p_0,\; p_2\; {\rm and}\; p_3$ satisfy the following constraints, 
\begin{equation}\label{cts}
\sum_{j=0,2,3}p_{j} = 1 \;\;\;\; , \;\;\;\; \sum_{j=0,2,3}p^2_{j} = 1.
\end{equation}
To find an explicit solution, one should either fix or interpret all the constants introduced in the above general solution, specially the essential ones \cite{Mac}, 
using different criteria. 
For example this is done in \cite{Bonn1}, in which the authors have used the above general form for negative $\Lambda$ as 
an exterior solution, matched to an infinite cylinder of dust cut out of a G\"{o}del universe as the interior solution, to form a globally regular solution.\\
Here, on the other hand, we show that one could fix all the above constants, to find  explicit 
solutions with $\Lambda$ as the only parameter of the stationary spacetime. To do so we employ just one criterion, and that is,
{\it the smooth reduction of the above general solution to the Minkowski spacetime in the absence of $\Lambda$}.\\
In what follows we apply the above criterion to find the stationary solution for $\Lambda > 0$. The negative $\Lambda$ solution is 
easily obtained following the same procedure.\\
Substituting $G_{0}(\rho)$ and $U(\rho)$ for $\Lambda > 0$ in \eqref{ds007}, we have,
\begin{align}\label{ds0071}
ds^2 &= (\frac{1}{a})^{2/3}\sin^{2/3}(a\rho)\cos^{2/3}(a\rho)\bigg([T_{1}^2 \tan^{2(p_{0}-1/3)U}(a\rho) - X_{1}^2 \tan^{2(p_{3}-1/3)}(a\rho)]dt^2 
\nonumber \\& - Z^2 \tan^{2(p_{2} - 1/3)}(a\rho)dz^2 
\nonumber \\&- [X_{2}^2 \tan^{2(p_{3}-1/3)}(a\rho) - 
T_{2}^2 \tan^{2(p_{0}-1/3)}(a\rho)]d{\phi}^2 \nonumber \\& - 
2[X_{1}X_{2} \tan^{2(p_{3}-1/3)}(a\rho)- T_{1}T_{2} \tan^{2(p_{0}-1/3)}(a\rho)]dt d\phi \bigg) - d\rho^2.  
\end{align}
in which $a \equiv \frac{\sqrt{3|\Lambda|}}{2}$. Now focusing on the $g_{zz}$ component and applying our criterion that it reduces to its flat space value ($g_{zz}=1$) in the limit $\Lambda \rightarrow 0$ ($a \rightarrow 0$),
sets $p_2=0$ and $Z^2 = a^{\frac{2}{3}}$. 
Using this value for $p_2$, and applying the constraints \eqref{cts} on $p_j$s, we immediately find that $p_0 = 0$ and $p_3 =1 $. Now focusing on the $g_{\phi\phi}$ component 
and require its reduction to the flat space value ($g_{\phi\phi} = \rho^2$), 
in the absence of $\Lambda$, sets $T_2=0$ and ${X_2} = (\frac{1}{a})^{2/3}$ 
\footnote{The same limit shows that a nonzero $T_2$ acts as a deficit angle parameter.}.
Now substituting the above selected values for $X_2$ and ${T_2}$ into \eqref{ds0071}, we end up with,
\begin{align}\label{ds008}
ds^2 &= (\frac{1}{a})^{2/3}\cos^{-2/3}(a\rho)[T_{1}^2 \cos^2 (a\rho)  - X_{1}^2 \sin^2 (a\rho)]dt^2 - d\rho^2 \nonumber \\& - \cos^{4/3}(a\rho)dz^2 
- (\frac{1}{a})^{2/3} \sin^2 (a\rho) \cos^{-2/3}(a\rho) [(\frac{1}{a})^{4/3} d{\phi}^2 -  2 X_{1}(\frac{1}{a})^{2/3}  d\phi dt].  
\end{align}
To have both $g_{tt} \rightarrow 1$ and $g_{t\phi} \rightarrow 0$, when $\Lambda \rightarrow 0$, we choose $X_{1}^2 = T_{1}^2 = a^{2/3}$, leading to the explicit solution, 
\begin{align}\label{ds009}
ds^2 &= \cos^{-2/3}(a\rho)\cos(2a\rho) dt^2 - d\rho^2 - \cos^{4/3}(a\rho)dz^2 \nonumber \\& 
-\frac{1}{a^2}\sin^2 (a\rho) \cos^{-2/3}(a\rho)d{\phi}^2  + \frac{2}{a} \sin^2 (a\rho) \cos^{-2/3}(a\rho)  dt d\phi,  
\end{align}
in which the radial coordinate is obviously restricted to the range $0 < \rho < \frac{\pi}{2\sqrt{3\Lambda}}$. The above spacetime is the stationary analog of the 
dS spacetime. Transformed into the $1+3$ form  of \eqref{ds06}, we find,
\begin{gather}\label{cyldes2}
 ds^2=\bigg(\cos^{-2/3}(a\rho) + \cos(2 a \rho)\bigg) \bigg(dt - \frac{\sin^2(a\rho)}{a \cos(2a\rho)}d\phi\bigg)^2  - d\rho^2 - \cos^{4/3}(a\rho) dz^2 \nonumber \\
 -\frac {1}{a^2} \frac{\sin^2(a\rho)\cos^{4/3}(a\rho)}{\cos(2a\rho)}d \phi^2.
\end{gather}
The static counterpart of \eqref{ds009}, which is one of the de Sitter-type spacetimes \cite{NKR}, is given by \cite{GP, Exact1},
\begin{equation}\label{cyldes}
 ds^2=\cos^{4/3} (a\rho) (d t^2
 - d z^2) - d\rho^2 - \frac {1}{a^2}
 \sin^2 (a\rho)
 \cos^{-2/3}(a\rho) d\phi^2,
\end{equation}
Obviously one could have obtained this line element from \eqref{ds008} by setting  $ X_1 = 0$ and $T_1 = 1 $.\\
Applying the above procedure to the case of general $\Lambda < 0$ solution , we end up with,
\begin{align}\label{ds0091}
ds^2 &= \cosh^{-2/3}(a\rho) dt^2 - d\rho^2 - \cosh^{4/3}(a\rho)dz^2 \nonumber \\& 
-\frac{1}{a^2}\sinh^2 (a\rho) \cosh^{-2/3}(a\rho)d{\phi}^2  + \frac{2}{a} \sinh^2 (a\rho) \cosh^{-2/3}(a\rho)  dt d\phi.  
\end{align}
in which now $0 < \rho < \infty$. This is obviously the stationary analog of the AdS spacetime. 
In the $1+3$ form  of \eqref{ds06}, it is given by,
\begin{gather}
ds^2=\cosh^{-2/3}(a\rho) \bigg(dt - \frac{\sinh^2(a\rho)}{a}d \phi \bigg)^2  - d\rho^2 - \cosh^{4/3}(a\rho) dz^2 \nonumber \\
-\frac {1}{a^2} \sinh^2(a\rho)\bigg( \cosh^{-2/3}(a\rho) + \sinh^2(a\rho) \bigg) d\phi^2,\label{ds0666}
\end{gather} 
The static counterpart of \eqref{ds0091} is given by \cite{Bonnor}, 
\footnote{Bonnor discovered this solution by setting the parameter related to the (linear) mass 
density equal to zero, in a static cylindrical solution given by Zofka and Bicak \cite{Zof}.}  
\begin{align}\label{ds0099}
 ds^2=\cosh^{4/3} (a\rho) (d t^2
 - d z^2) - d\rho^2 - \frac {1}{a^2}
 \sinh^2 (a\rho)
 \cosh^{-2/3}(a\rho) d\phi^2.
\end{align}
To the best of our knowledge, the above {\it explicit} forms of stationary analogs of dS and AdS
spacestimes, \eqref{ds009} and \eqref{ds0091}, are introduced here for the first time in the literature 
\footnote{Indeed even in the main exact solution book \cite {Exact1}, in section 22.2 (stationary cylindrically-symmetric fields), in the subsection on ``$\Lambda$-term solutions'' 
the authors have given only the general static family, Eq. (2.9) of \cite{Mac}, instead of the general stationary family \eqref{ds007}, which is the Eq. (2.16) of \cite{Mac}.}.\\
It is worthwhile to mention some of the features of the above stationary solutions and compare them with those of dS and AdS spacetimes:\\
I-They are both regular flat on the axis, $\rho = 0$, similar to  dS and AdS which are regular flat at $r=0$.\\
II-The stationary analog of the dS spacetime, \eqref{ds009}, like de-Sitter space has an admissible range of the radial coordinate $0 < \rho < \frac{\pi}{2\sqrt{3\Lambda}}$, 
while its negative $\Lambda$ counterpart, \eqref{ds0091}, like AdS space is valid for the whole range of the radial coordinate.\\
III-These solutions, like static dS and AdS spacetimes, are in non-comoving coordinate systems which is evident from 
their non-zero Christoffel symbol $\Gamma^\rho_{tt}$.\\
Regularity of the above stationary solutions on the axis, is also evident in their Kretschmann invariants 
\begin{gather}
K=\frac{4\Lambda^2}{3} \big(2+ \sec^4 (a\rho)\big) \;\;\;\;\;\; \Lambda > 0 \\ 
K=\frac{4\Lambda^2}{3} \big(2+ \sech^4(a\rho)\big) \;\;\;\;\;\; \Lambda < 0,
\end{gather} 
which interestingly enough, they share with their static counterparts, and both reduce to $K = 4\Lambda^2$ on the axis.
\section{The cosmological term as a dark fluid}
To see why one should consider a dark fluid nature for the cosmological term or otherwise deal with a dilemma, we scrutinize, as we did in the case of
the de Sitter-type spacetimes \cite{NKR}, the anisotropic nature of the above stationary gravitational fields. First of all, without the
detailed calculation of the geodesics, one could see that a test particle released near the symmetry axis in the stationary analog of dS space \eqref{ds009}, 
will start rotating around the axis in the $z=constant$
hypersurface with an increasing radius. This could be easily understood in terms of the garvitoelectromagnetic 
fields of the solution \eqref{ds009} which are given by \cite{MN,LL},
\begin{gather}\label{E&B}
E^\rho_g = \frac{\sqrt{3\Lambda}}{12} \tan ^2({\sqrt{3\Lambda}}\rho) \big(5 + \tan ^2(\frac{\sqrt{3\Lambda}}{2}\rho)\big) \\
{B^z_g} = 2\sqrt{3\Lambda}\cos^{-1/2}(\sqrt{3\Lambda}\rho) \cos^{-1/3}(\frac{\sqrt{3\Lambda}}{2}\rho).
\end{gather}
They contribute to the gravitational (Lorentz) 3-force, 
${\bf f}_g =  \frac{m}{\sqrt{1-{v^2}}}\left( {\bf E}_g + {\rm {\bf v}}\times \sqrt{g_{00}}{\bf B}_g\right)$,
which acts on the test particles 
\footnote{Employing the quasi-Maxwell form of the Einstein field equations with  multi-component perfect fluid sources, 
it is shown in \cite{MAN}, that one could introduce a physical characterization of the well-known static and stationary perfect fluid solutions
according to their gravitoelectric and gravitomagnetic fields.}. 
Also by the same token, one can show that the above gravitomagnetic field $B_g$ is the origin of the rotation of the
local inertial frames as well as the Lense-Thirring effect \cite{Rindlerb}, and the related effects such as the gravitomagnetic clock effect \cite{Mashhoon}.
In other words the stationary analog of the  dS spacetime, produces an {\it anisotropic} field reminiscent of fields produced by rotating sources. 
To clarify more on this point, comparison of \eqref{ds0666} with the G\"{o}del Universe, which is also a stationary cylindrically symmetric spacetime, is in order. 
The G\"{o}del universe in cylindrical coordinates is given by the following line element
\begin{equation}\label{ds033}
d{s^2} = \big({dt} - \frac {2\sqrt{2}}{\sqrt{-2\Lambda}} \sinh^2 (\frac{\sqrt{-2\Lambda}}{2}\rho) d\phi\big)^2 - d\rho^2 - dz^2 -
(\frac{1}{-2\Lambda}) \sinh^2(\sqrt{-2\Lambda}\rho) d\phi^2.
\end{equation}
It should be noted that the G\"{o}del universe is {\it not} an Einstein space, and two different energy-momentum tensors could be assigned 
to it as its source. Either a stiff matter ($p=\rho$), or  
a dust ($p=0$) plus negative cosmological constant, where $\Lambda = - 4\pi \rho_{dust}$. In both cases the matter component 
is in the {\it comoving frame} with 4-velocity $u^a = (\sqrt{-2\Lambda}, 0, 0, 0)$. Obviously the 
rotational (anisotropic) effects of the G\"{o}del universe could be attributed (traced back) to its geodesic {\it matter current} which rotates with constant velocity 
$\sqrt{-\Lambda} = \sqrt{4\pi \rho_{dust}}$ \cite{Rindlerb}.\\
Comparison of the negative $\Lambda$ solution, \eqref{ds0666} with the G\"{o}del universe shows that, although unlike G\"{o}del it does not contain closed timelike curves (CTCs),
the two solutions share the same regular flat space behavior near the axis ($\rho \rightarrow 0$), 
and more importantly, the similar structure of their gravitomagnetic potential leads to a gravitomagnetic field along the symmetry axis.
On the other hand unlike the G\"{o}del universe, both \eqref{ds009} and \eqref{ds0091} are matter-free solutions, and hence the 
anisotropic behavior in these solutions could only be assigned to their single parameter $\Lambda$, which is dynamic-free if is taken to be just a universal geometric constant. 
This indicates (as in the case of their static counterparts) that we could justify the anisotropy of these solutions, only if we think of the cosmological term as a 
dark fluid assigned with a {\it 4-velocity}, whose role is eclipsed in the non-comoving frame. This is simply due to the fact that 
by setting $p=-\rho$ in the energy momentum tensor of a perfect (dark) 
fluid, the role of its 4-velocity becomes hidden both in the energy-momentum tensor of the fluid, and  
in the corresponding cosmological solutions. However in the above stationary analogs of dS and AdS spaces, unlike the case of the de Sitter-type spacetimes, 
we do not need to transform into the comoving frame to 
recover the role of the dark fluid's velocity in justifying their anisotropic expansion \cite{NKR}, and the stationarity of the solutions itself does the trick.
\section{Final note}
In the last two decades we have witnessed great interest in de Sitter and anti-de Sitter spacetimes in different contexts, both in theoretical 
physics and in cosmology. This is rooted in the two discoveries at the end of the 20th century, namely the introduction of the AdS/CFT correspondence, 
and the observation of the accelerated expansion of the Universe. It is therefore interesting to discover that there are 
stationary analogs of these spacetimes and they might be of somewhat similar application and use. Already as an interesting application of these solutions, 
here we have shown how their inherent anisotropy infer a dark fluid nature for the cosmological term.
\section *{Acknowledgments}
The authors would like to thank B. Mashhoon for careful reading of the manuscript and useful comments. M. N-Z thanks University of Tehran for supporting 
this project under the grants provided by the research council.


\end{document}